\documentclass[preprint]{aastex}

\received{}
\accepted{}
\journalid{}{}
\articleid{}{}
\slugcomment{}

\newcommand{\etal}{{\em et~al.}}
\newcommand{\ug}{$u'\!-\!g'$}
\newcommand{\gr}{$g'\!-\!r'$}
\newcommand{\ri}{$r'\!-\!i'$}
\newcommand{\iz}{$i'\!-\!z'$}
\newcommand{\ugriz}{$u',g',r',i',z'$}

\shortauthors{Budav\'ari et~al.}
\shorttitle{Photometric redshifts from reconstructed QSO templates}

\begin{document}

\title{Photometric redshifts from reconstructed QSO templates}

\author{Tam\'as  Budav\'ari\altaffilmark{\ref{Eotvos},\ref{JHU}}, 
Istv\'an Csabai\altaffilmark{\ref{Eotvos},\ref{JHU}},
Alexander S. Szalay\altaffilmark{\ref{JHU}},
Andrew J. Connolly\altaffilmark{\ref{Pittsburgh}},
Gyula P. Szokoly\altaffilmark{\ref{AIP}},
Daniel E. Vanden Berk\altaffilmark{\ref{Fermilab}},
Gordon T. Richards\altaffilmark{\ref{PennState}},
Michael A. Weinstein\altaffilmark{\ref{PennState}},
Donald P. Schneider\altaffilmark{\ref{PennState}},
Narciso Ben\'{\i}tez\altaffilmark{\ref{JHU}},
J. Brinkmann\altaffilmark{\ref{APO}},
Robert Brunner\altaffilmark{\ref{Caltech}},
Patrick B. Hall\altaffilmark{\ref{Princeton},\ref{Chile}},
Greg Hennessy\altaffilmark{\ref{USNO}},
\v{Z}eljko Ivezi\'{c}\altaffilmark{\ref{Princeton}},
P\'eter Z. Kunszt\altaffilmark{\ref{JHU}},
Jeffrey A. Munn\altaffilmark{\ref{USNO}},
R. C. Nichol\altaffilmark{\ref{CMU}},
Jeffrey R. Pier\altaffilmark{\ref{USNO}},
and
Donald G. York\altaffilmark{\ref{Chicago}}
}

\altaffiltext{1}{Department of Physics, E\"{o}tv\"{o}s University,
Budapest, Pf.\ 32, Hungary, H-1518
\label{Eotvos}}

\altaffiltext{2}{Department of Physics and Astronomy, The Johns Hopkins University, 3701 San Martin Drive, Baltimore, MD~21218
\label{JHU}}

\altaffiltext{3}{Department of Physics and Astronomy, University of
Pittsburgh, Pittsburgh, PA 15260
\label{Pittsburgh}}

\altaffiltext{4}{Astrophysikalisches Institut Potsdam, An der
Sternwarte 16, D-14482 Potsdam, Germany 
\label{AIP}}

\altaffiltext{5}{Fermi National Accelerator Laboratory, P.O. Box 500,
Batavia, IL~60510
\label{Fermilab}}

\altaffiltext{6}{Department of Astronomy and Astrophysics,
The Pennsylvania State University, University Park, PA 16802
\label{PennState}}

\altaffiltext{7}{Apache Point Observatory, P.O. Box 59,
Sunspot, NM 88349-0059
\label{APO}}

\altaffiltext{8}{Department of Astronomy, California Institute of
Technology, Pasadena, CA 91125 
\label{Caltech}}

\altaffiltext{9}{Princeton University Observatory, Princeton, NJ~08544
\label{Princeton}}

\altaffiltext{10}{Pontificia Universidad Cat\'{o}lica de Chile,
Departamento de Astronom\'{\i}a y Astrof\'{\i}sica,
Facultad de F\'{\i}sica, Casilla 306, Santiago 22, Chile
\label{Chile}}

\altaffiltext{11}{U.S. Naval Observatory, 
3450 Massachusetts Ave., NW, 
Washington, DC  20392-5420
\label{USNO}}

\altaffiltext{12}{Dept. of Physics, Carnegie Mellon University,
5000 Forbes Ave., Pittsburgh, PA~15232
\label{CMU}}

\altaffiltext{13}{University of Chicago, Astronomy \& Astrophysics
Center, 5640 S. Ellis Ave., Chicago, IL 60637
\label{Chicago}}

\begin{abstract}
From SDSS commissioning photometric and spectroscopic data, we investigate
the utility of photometric redshift techniques to the task of estimating QSO
redshifts. We consider empirical methods (e.g.\ nearest-neighbor searches and
polynomial fitting), standard spectral template fitting and hybrid approaches
(i.e.\ training spectral templates from spectroscopic and photometric
observations of QSOs). We find that in all cases, due to the presence of
strong emission-lines within the QSO spectra, the nearest-neighbor and
template fitting methods are superior to the polynomial fitting
approach. Applying a novel reconstruction technique, we can, from the SDSS
multicolor photometry, reconstruct a statistical representation of the
underlying SEDs of the SDSS QSOs. Although, the reconstructed templates are
based on only broadband photometry the common emission lines present within
the QSO spectra can be recovered in the resulting spectral energy
distributions.  The technique should be useful in searching for spectral
differences among QSOs at a given redshift, in searching for spectral
evolution of QSOs, in comparing photometric redshifts for objects beyond the
SDSS spectroscopic sample with those in the well calibrated photometric
redshifts for objects brighter than 20th magnitude and in searching for
systematic and time variable effects in the SDSS broad band photometric and
spectral photometric calibrations.
\end{abstract}


\section{Introduction} \label{sec:intro}

In recent years, with new multicolor surveys coming on-line, the application
of photometric redshifts to the analysis of the physical properties of
galaxies has become increasingly popular, (see \citet{weymann99} for recent
works).  Photometric redshift estimation techniques rely on detecting the
passage of continuum features within the spectral energy distribution (SED)
of sources (e.g.\ for galaxies, the 4000{\AA} break) across a series of
photometric passbands. For objects with featureless spectra, the photometric
redshift estimation is extremely difficult or impossible as the colors of
these objects vary slowly (if at all) with redshift. Emission lines may,
however, help if they are strong enough relative to the continuum flux to be
detectable within the broadband photometry.  As quasar emission lines carry a
significant amount of flux \citep{francis91, richards01a}, we expect that
they will provide sufficient leverage to help in the redshift prediction for
lower redshift objects.  In addition, it may be easier to estimate redshifts
for high-redshift objects ($z \gtrsim 3$) due to the blanketing of the
Ly\,$\alpha$ forest (e.g.\ \citet{madau95}).  In the astronomical
literature, there are a number of different approaches for estimating
photometric redshifts.  Most techniques can be categorized into two basic
classes; those which use spectral energy distributions as spectral templates
derived from models or observations \citep{koo85, gwyn96, sawicki97,
connolly99, soto99, benitez00, budavari99, budavari00, csabai00} and those
which establish a direct empirical relation between colors and redshift
\citep{connolly95a, wang98, brunner99} using a training set.

In this paper and the companion paper by \citet{richards01b}, we investigate
these techniques and determine how successfully they may be applied to the
question of the estimation of the redshifts of QSOs.  In this study, we put
more emphasis on the so-called template fitting method and quasar template
reconstruction.  
In the companion paper, we discuss an empirical method that
essentially assumes that quasar colors are alike at a given redshift. 
We use a set of $\sim 2600$ known QSOs with five band
\citep[\ugriz;][]{fukugita,lupton} Sloan Digital Sky Survey \citep{york}
photometry and redshifts in the $0<z<5$ range.
In the \citet{richards01b}
paper, we study the effects caused by reddened quasars and extended objects
concerning redshift outliers, and also present more examples of science that
will benefit from QSO photometric redshifts. Observational and target
selection details are also given there.

In Section~\ref{sec:emp}, this study considers empirical methods to estimate
redshift. In Section~\ref{sec:temp}, we apply the standard template fitting
technique using composite spectra, and in Section~\ref{sec:tempRec} and
\ref{sec:specTypes}, we extend our analysis to include SEDs reconstructed
from photometric information.

\section{Empirical approach} \label{sec:emp}

The simplest way of getting photometric redshifts is empirical fitting.  We
assume the redshift, $z$, to be a simple function of the magnitudes or
colors.  \citet{connolly95a} used low order polynomial functions for
galaxies.  The fitting formula (redshift vs.\ colors) is derived by using a
training set of photometric data {\em and} spectroscopic redshifts.  After
the calibration, this simple analytic function can predict redshifts for
objects directly from photometric observations. While this approach has
proven successful for studies of the photometric redshifts of galaxies, the
spectral energy distributions of quasars differ from galaxies in many ways.
In the optical and ultraviolet, a quasar spectrum consists of an
approximately power law continuum together with strong emission and
absorption lines. Thus quasar colors can change drastically over the space of
a small interval in redshift (as an emission line passes from one passband to
the next) or can remain constant with redshift due to the power-law
dependency of the continuum.  Given this situation, we would not expect
polynomial fitting to yield accurate and reliable photometric redshifts for
QSOs due to the fact that the polynomial functions used vary slowly with
redshift. This preconception is borne out by the SDSS data. Fitting both
quadratic and cubic relations between the colors of the quasars and their
redshifts gave almost no correlation between prediction and spectroscopic
redshifts.

A more adaptive technique than polynomial fitting is that of nearest neighbor
(NN) mapping. This technique assigns the redshift of a source to that of the
closest object in color space (with a known redshift). This approach is,
essentially, a piecewise constant function.  Although, as we stated above,
the polynomial technique does not work, a combination of the polynomial
technique and the NN fitting technique can also be applied where we find the
nearest neighbors and locally fit the polynomial relation over objects within
a given radius.  This method can suppress any unwanted effect due to possible
erroneous nearest neighbors, but would require more data.  
A similar empirical method is developed in the companion paper
\citep{richards01b}, that implements an NN estimator where the reference
points are derived from colors averaged over redshift bins.

\begin{figure}
\epsscale{0.85}
\plotone{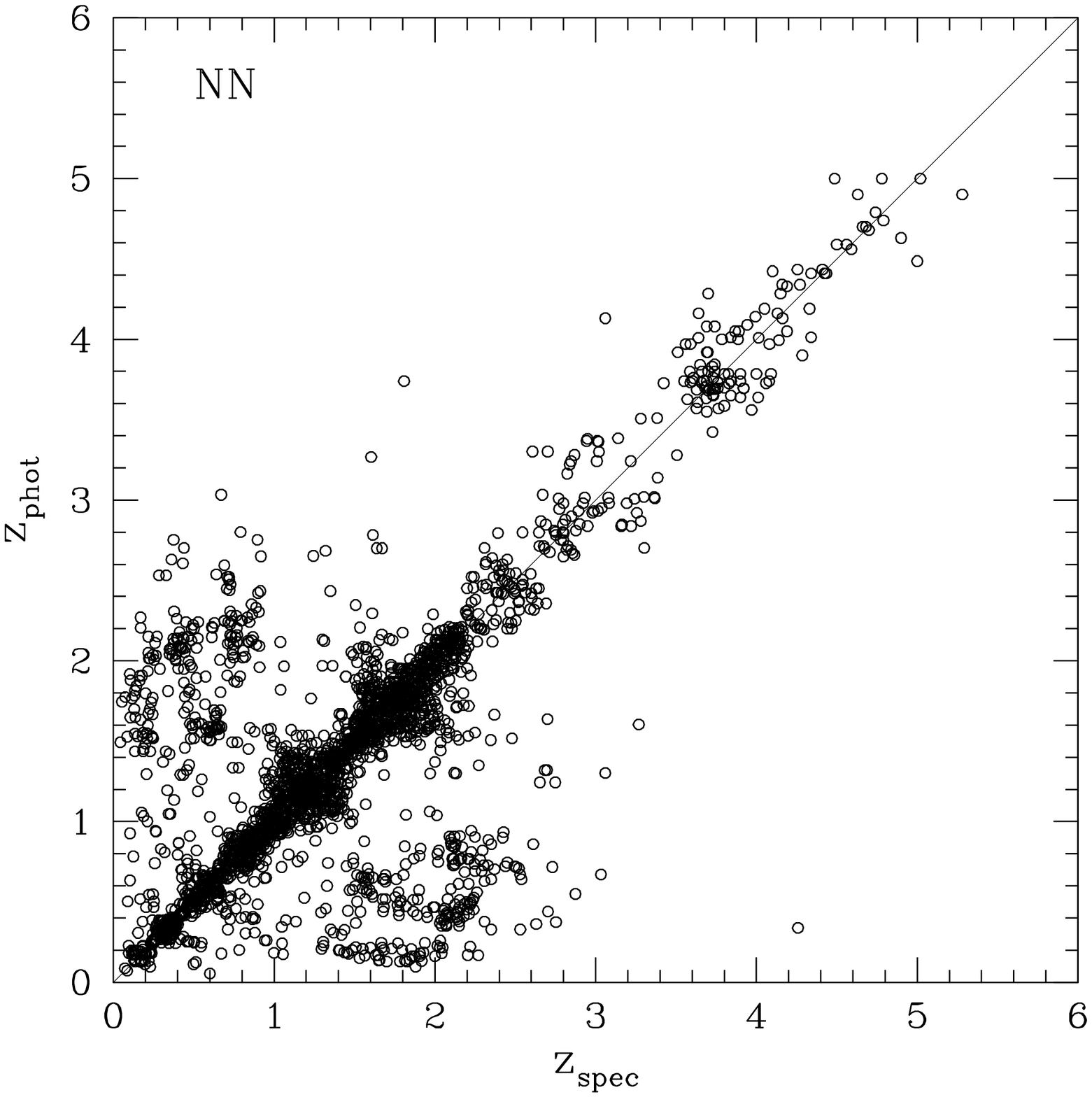}
\caption{Photometric redshifts from the  nearest neighbor estimator. The bulk
of the objects have accurate  redshifts without any systematic deviation, but
the photometric degeneracy  confuses quite a few estimates.  The rms error is
$\Delta_{\rm all}=0.64$  for all  objects and $\Delta_{\rm  0.3}=0.116$ when
excluding outliers (see text).
\label{fig:nn}}
\end{figure}

Figure~\ref{fig:nn} shows the correlation between the photometric and
spectroscopic redshifts from applying the NN estimator.
Most of the predicted redshifts ($\sim$70\%) match well the
spectroscopic measurements. There are, however, a significant number of
outliers (i.e.\ the distribution of the prediction error is clearly
non-Gaussian).  This is due to the color degeneracy, i.e.\ within the
photometric uncertainties identical objects (in colors \ug, \gr, \ri, \iz)
have significantly different spectroscopic redshifts.  For this reason, for
this figure and for all of the following spectroscopic redshift ($z_{\rm
spec}$) vs.\ photometric redshift ($z_{\rm phot}$) comparison figures we
calculate two rms values: one for all of the objects ($\Delta_{\rm all}$) and
one for the objects that have $|z_{\rm spec} - z_{\rm phot}| < 0.3$
($\Delta_{0.3}$). The former of these measures is used to estimate the effect
of the contamination of a QSO photometric-redshift sample by catastrophic
failures and the latter statistic is to measure the intrinsic accuracy of the
photometric redshift relation for different analysis techniques. For the NN
estimator these values are $\Delta_{\rm all} = 0.64$ and $\Delta_{0.3} =
0.116$. Because of the degeneracy, the number of outliers ($\Delta z > 0.3$)
is quite large, approximately 30\%.
  We also note that these results are for a sample of objects already known
to be quasars.  If the input sample includes objects that turn out not to be
quasars, then this will dilute our results.

To determine whether the magnitude of a QSO might alleviate the degeneracy
between the colors of QSOs as a function of redshift, we applied the nearest
neighbor method in magnitude space. We find that this degeneracy is also
present when using magnitudes.  One hope for removing this effect may come
from extending the photometric observations to longer or shorter wavelengths,
e.g.\ using UV or IR data.  Plausible reasons for the large number of
redshift outliers are discussed in \citet{richards01b}.  We note that the
behavior of the estimation error seems to change around $z \sim 3$ with the
outliers almost completely disappearing.  This is the redshift where the
Lyman-break passes out of the $u'$ filter, which introduces a strong
continuum feature into the quasar SED, and at higher redshifts, the continuum
blanketing of the Ly\,$\alpha$ forest has even more significant effect on the
observed spectrum.  This attenuation enhances the quality of the redshift
estimation (as is the case for high redshift galaxies). It is also important
to note that these empirical methods seriously break down outside the
redshift range that the training set is in. Extrapolation is impossible, for
example, to higher redshifts. The method also assumes that the set of objects
used for training is representative of the objects studied.  Finally,
transformation to other filter systems is extremely hard, if not impossible,
without using spectral templates.

\section{Template fitting} \label{sec:temp}

Template fitting photometric-redshift methods have the advantage of not
requiring a training set (we assume that the physics of the QSO SEDs is fully
encoded within our distribution of templates). We do, however, require a
complete spectral library that has spectral templates that cover all of the
spectral types of sources within our sample and that extend over the
restframe wavelength ranges that our photometric observations occupy (for
high and low redshift sources). In its simplest form this template library
could be the mean spectrum of a sample of SEDs. For galaxies this is not a
realistic option since the continuum shape of different spectral types of
galaxies varies drastically.  For quasars, there are no well-defined classes
of continuum spectral features, mostly because previous observations
have indicated that quasars have SEDs with similar power law continua.

In an attempt to construct such a composite mean QSO spectrum, we consider
those available within the astronomical literature.  We tested the method
with the LBQS composite \citep{francis91} spectrum. In the meantime the SDSS
composite spectrum \citep{vanden01} became available; in this study, we
present results using this composite.  In Figure~\ref{fig:zz1} we plot
photometric redshifts from template fitting against spectroscopic redshifts,
where template colors are derived by convolving the SDSS filter functions
with the SDSS composite spectrum as well as by convolving the filter
functions with our reconstructed template (Figure~\ref{fig:composite}; see
later).  The figure is similar to that found in Figure~\ref{fig:nn}: the
outliers are more stratified in the template fitting case and the dispersion
in photometric vs.\ spectroscopic redshift is somewhat larger than in the NN
case ($\Delta_{\rm all} = 0.89$ and $\Delta_{0.3} = 0.123$).  Of course
direct empirical methods give less astrophysical information --- as opposed
to template fitting that provides consistent spectral type and luminosity
besides redshift --- and are hard to apply to different sets of
observations, e.g.\ a sample with a different filter set.

The template fitting estimator does not seem to be superior when compared to
NN estimators (see above and also \citet{richards01b}) if we use a single
composite SED. In some sense it can be considered worse; there are
approximately 20\% more outliers.  It is likely, therefore, that the
composite template assumed for the photometric redshift estimation is not
representative of the full QSO sample, and that we might need to introduce
additional quasar spectral types (e.g.\ corresponding to BALs, Fe\,{\sc ii}
emission objects, etc.)\ to our analysis. 
  Empirical estimators, like the NN method or the one
discussed in the companion paper, do not assume that QSOs at all redshifts have
similar restframe spectra as opposed to the template fitting technique
using a composite SED. 
Thus the fact that the empirical
estimators yield better redshifts than a single composite may also imply that
restframe QSO spectra correlate with redshift.  We explore these ideas in the
following section.

\section{Template reconstruction} \label{sec:tempRec}

Without having access to a complete quasar template library, we adopt a
hybrid method developed by \citet{budavari99, budavari00} and
\citet{csabai00} to bring together the advantages of empirical and template
fitting techniques.  This method requires a training set of objects with
spectroscopic redshifts and multicolor photometry to establish a statistical
representation of the underlying QSO SEDs. The templates are created to yield
simulated colors in the best possible agreement with the photometry, and
using them for photometric redshift estimation is just one application.  The
method is also suitable for detecting possible differences between the
spectroscopic and photometric flux calibrations and to detect whether the
introduction of different QSO spectral types are needed.

We propose a novel technique, similar to what is known as learning vector
quantization \citep{kohonen} in the neural network literature, that
reconstructs discrete SEDs in an iterative way.  The procedure called {\em
adaptive spectrum quantization} (hereafter ASQ) is so robust that it can
develop a set of template spectra from scratch, i.e.\ from one single
constant function.
The idea here is to separate quasars into coherent classes of spectral
templates based on their photometry and spectroscopic redshifts. The
algorithm improves the templates step by step to increase the agreement with
the photometric observations. Since the method is capable of adding and
removing classes, it can be started from a single class.  One loop of our
iterative procedure consists of four steps:
\begin{itemize}
\item[{\it a.})] All objects in the  training set are categorized into
classes of the most likely templates. 
\item[{\it b.})] The  estimated SEDs of the objects are {\em repaired}.   
\item[{\it c.})] The reference templates are replaced with the mean of the
repaired spectra.
\item[{\it d.})] The templates are dynamically duplicated or removed if
statistically desired.  
\end{itemize}

Let $\Psi=\left\{\psi_i(\lambda)\right\}_{i=1}^N$ represent the initial set
of spectral templates, where $N$ is the number of templates and $i=1, ..., N$.
Given the spectroscopic redshift and multicolor photometry of an object, the
most likely template can be selected from a $\chi^2$
optimization. Unfortunately, in general, even the best fitting template
$\psi_k(\lambda)$ is not perfect in the sense that it cannot reproduce the
measured colors if convolved with the throughput of the filters and
instrument response. So the $\psi_k(\lambda)$ spectrum must be ``repaired''
according to the observational constraints (i.e.\ the multicolor photometry)
as described in \citet{budavari00}. The adjusted spectrum
\begin{equation}
\psi_k'(\lambda) = \psi_k(\lambda) + \delta\psi_k(\lambda)
\end{equation}
will be in better agreement with the measurements, but may be affected by some
irregularity of the object.  So instead of replacing the original
$\psi_k(\lambda)$ with $\psi_k'(\lambda)$ or adding the repaired SED to the
original $\Psi$ set, one can collect all the adjusted spectra assigned to the
$k$th template and replace $\psi_k(\lambda)$ with the average of these
repaired template spectra. A similar reparation can be applied to all
templates used in the analysis.
\begin{equation}
\psi_k(\lambda) \to
	\langle \psi_k'(\lambda) \rangle
 	= \psi_k(\lambda) + \langle \delta\psi_k(\lambda) \rangle
\end{equation}

Introducing additional reference templates requires more deliberation.  One
can think of several conditions when a class of objects should be split into
two (e.g.\ based on the number of assigned objects etc.)\ but when the
observed training set is limited in size, the choice of this condition may be
crucial.  When the iteration converges with a fixed number of templates
(relaxation), a new reference spectral template can be added to the set if
deemed necessary due to a large intrinsic scatter around the mean repaired
spectrum $\langle \psi_k'(\lambda) \rangle$ or due to an excess of member
objects within a particular spectral class. To split a class one can make a
copy of the corresponding template and perturb it slightly by shifting its
spectral shape towards that of the modulated spectrum of a random member
within its spectral class.  A more elaborate discussion of the method will be
described elsewhere \citep{budavari01}.

\begin{figure}
\epsscale{0.85}
\plotone{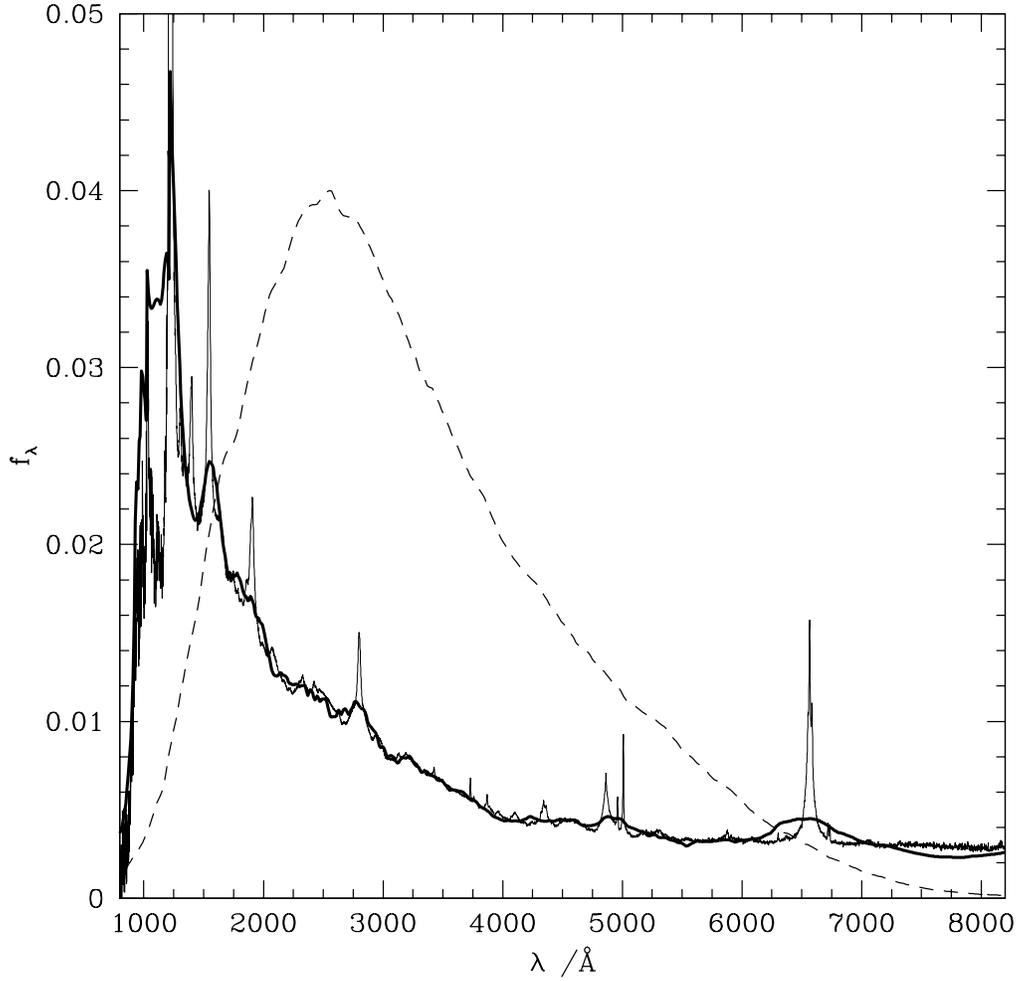}
\caption{Comparison of composite SEDs.  The SDSS composite QSO spectrum (thin
line) and an average SED derived from photometry and redshifts of SDSS quasar
sample are seen (thick line).  The power-law continuum shape of the trained
SED tracks nicely the thin line and also the most powerful spectral features
(e.g.\ Ly\,$\alpha$, C\,{\sc iv}, Mg\,{\sc ii}) are seen. The dashed line is
the scaled constraint curve (see text for details). 
\label{fig:composite}}
\end{figure}

We tested the above method for galaxies --- where the photometric redshift
estimation techniques are well established --- and it provides photometric
redshifts for galaxies, that are in good agreement with those estimates based
on eigentemplates.  Quasar redshifts can also be predicted with
accuracy. Starting from a featureless constant SED, we develop an average
template that is very similar to the empirical SDSS composite spectrum
\citep{vanden01} as seen in Figure~\ref{fig:composite}.  Our reconstructed
template (derived purely from the photometric data) tracks the power-law
continuum of the high-resolution spectrum to a remarkable degree of accuracy.
The good correspondence between the continuum shapes shows not only the
strength of the reconstruction algorithm, but also that the efforts the SDSS
team made for the correct spectrophotometric calibration were fruitful.

The strongest spectral lines, such as Ly\,$\alpha$ (at 1216\AA, merged with
N\,{\sc v} at 1240{\AA}), C\,{\sc iv} (1548\AA), Mg\,{\sc ii} (2798\AA) are
clearly visible. The H\,$\beta$ (4861\AA) and the [O\,{\sc iii}] (4959\AA,
5007\AA) lines are merged into a single feature and the blurred H\,$\alpha$ (at
6563\AA) line is also present.  The sharpness of these lines is quite
surprising, given that only the broadband photometric information (a typical
filter width is 1500{\AA}) and the value of the spectroscopic redshift were
used as input parameters in the reconstruction algorithm.

For a given redshift, the photometric observation gives constraints on the
possible underlying SED, since we expect to get back the measured photometric
values by redshifting the SED and convolving it with the filter response
function.  This constraint obviously depends on the photometric system, and
also the redshift of the object as the restframe spectrum is sampled at
different wavelengths.  Adding up the filter curves, shifted to the
restframe, for all objects provides a smooth curve as a function of
wavelength, shown with dashed line in Figure~\ref{fig:composite}, that is
proportional to the amount of information available for the determination of
the shape of the composite. This is similar to what is typically shown for
composites derived from spectrophotometric observations, but computed from
the shape of the photometric transmission curves.  The more objects at
different redshifts we use to constrain the spectrum, the finer details could
be recovered, e.g.\ the smaller can be the width of the reconstructed line.
An additional effect is that short wavelength lines are constrained by higher
redshift objects so that when they are transformed back to the restframe
their linewidths get narrower. In Figure~\ref{fig:composite} one can see that
reconstructed lines at shorter wavelength are narrower than the ones at
longer wavelength.

\begin{figure}
\epsscale{0.99}
\plottwo{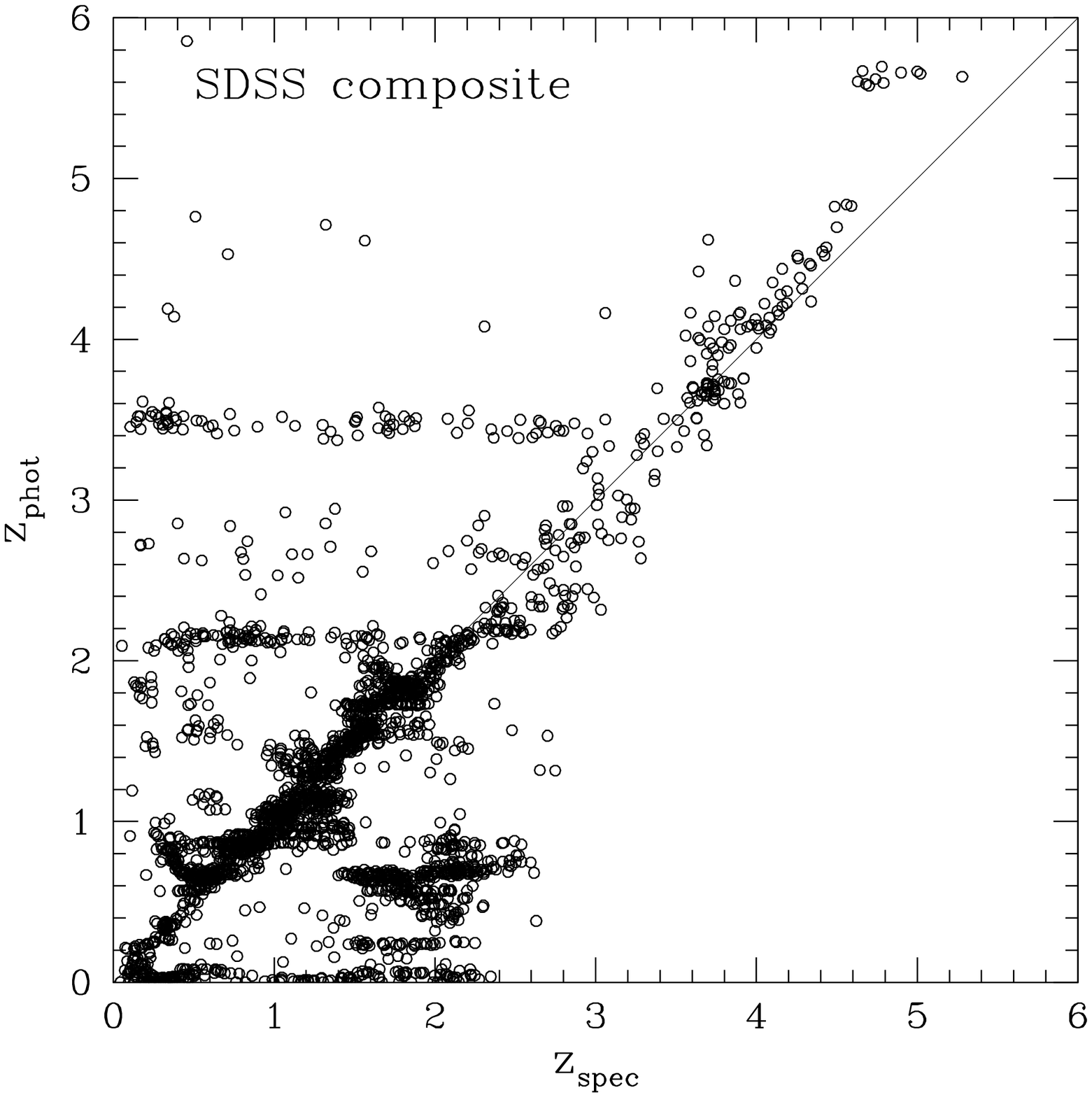}{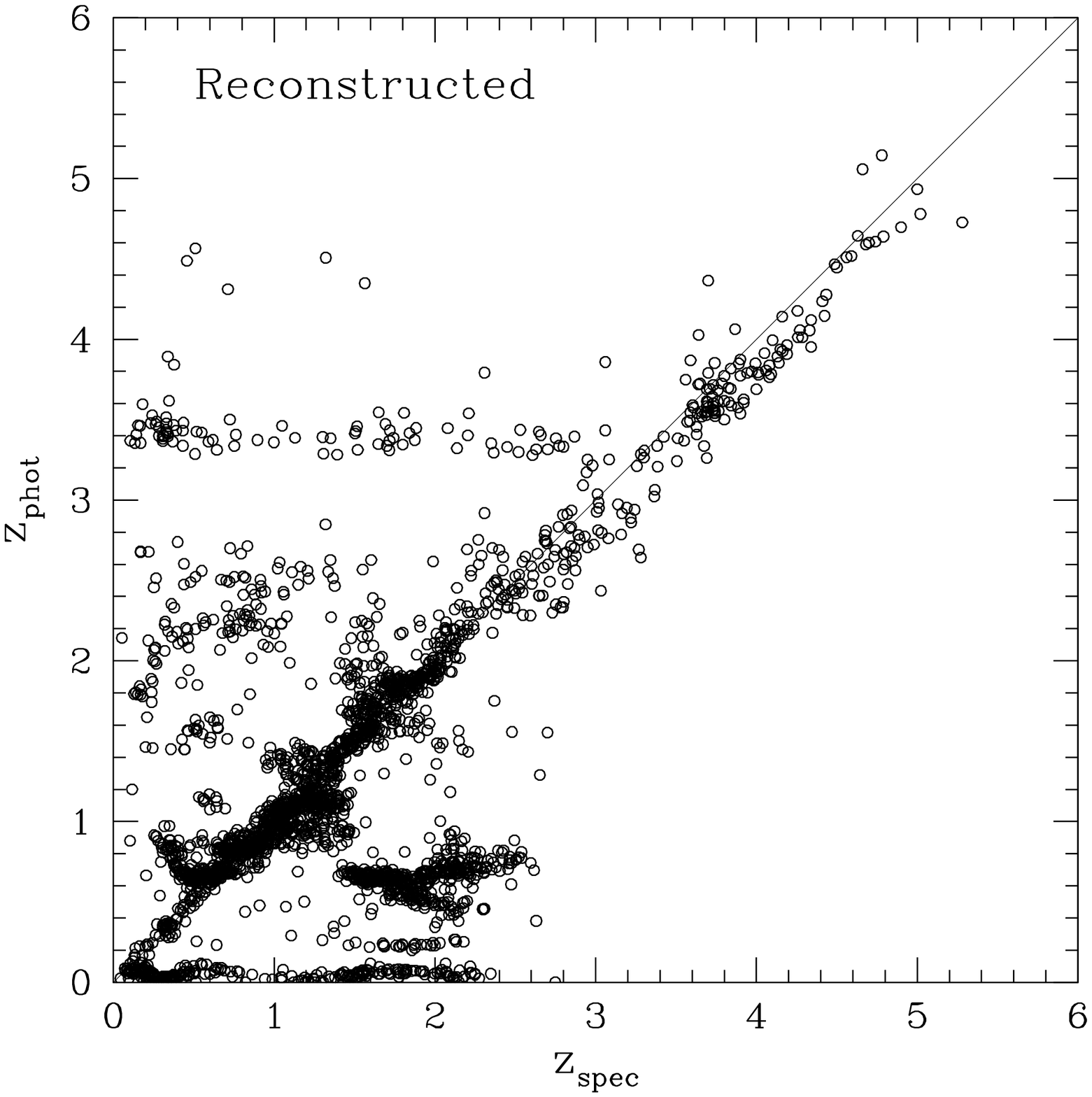}
\caption{Photometric redshifts from one  single SED. The SDSS composite based
estimates  are  seen  in  the  left  (rms  errors  $\Delta_{\rm  all}=0.84$,
$\Delta_{0.3}=0.128$)  and results  from  the reconstructed  template in  the
right panel ($\Delta_{\rm all}=0.89$, $\Delta_{0.3}=0.123$).
\label{fig:zz1}}
\end{figure}

The difference between the composite and reconstructed spectrum below
$\sim 1200${\AA} is due to the fact that the composite spectrum is not
corrected for inter-galactic attenuation.  Comparing the photometric
redshift estimations with the composite and the reconstructed spectrum
in Figures~\ref{fig:zz1}a and \ref{fig:zz1}b, we find little difference
except at high redshifts which is due to the effect of a different
handling of the inter-galactic attenuation.

\section{QSO spectral types} \label{sec:specTypes}

Now we arrive at the last step of the ASQ algorithm where we consider
whether there is evidence for more than just one spectral template.
The optimal number of templates is partly determined by the size of
the training set. If we divide a sample into too many subsets, there
will not be sufficient objects within any one subset to constrain the
reconstructed SEDs.  On the other hand, the more spectral types we
allow, the closer will be our resulting template library to the SEDs of
the quasars for which we wish to predict a redshift. Our experiments
show that, for the current training set, a QSO template library with 4
templates is the optimal choice. Introducing further templates does
not improve the agreement of synthetic and measured colors
significantly ($\chi^2$ decreases less than 10\%) and the photometric
redshifts do not become significantly better.

\begin{figure}
\epsscale{0.85}
\plotone{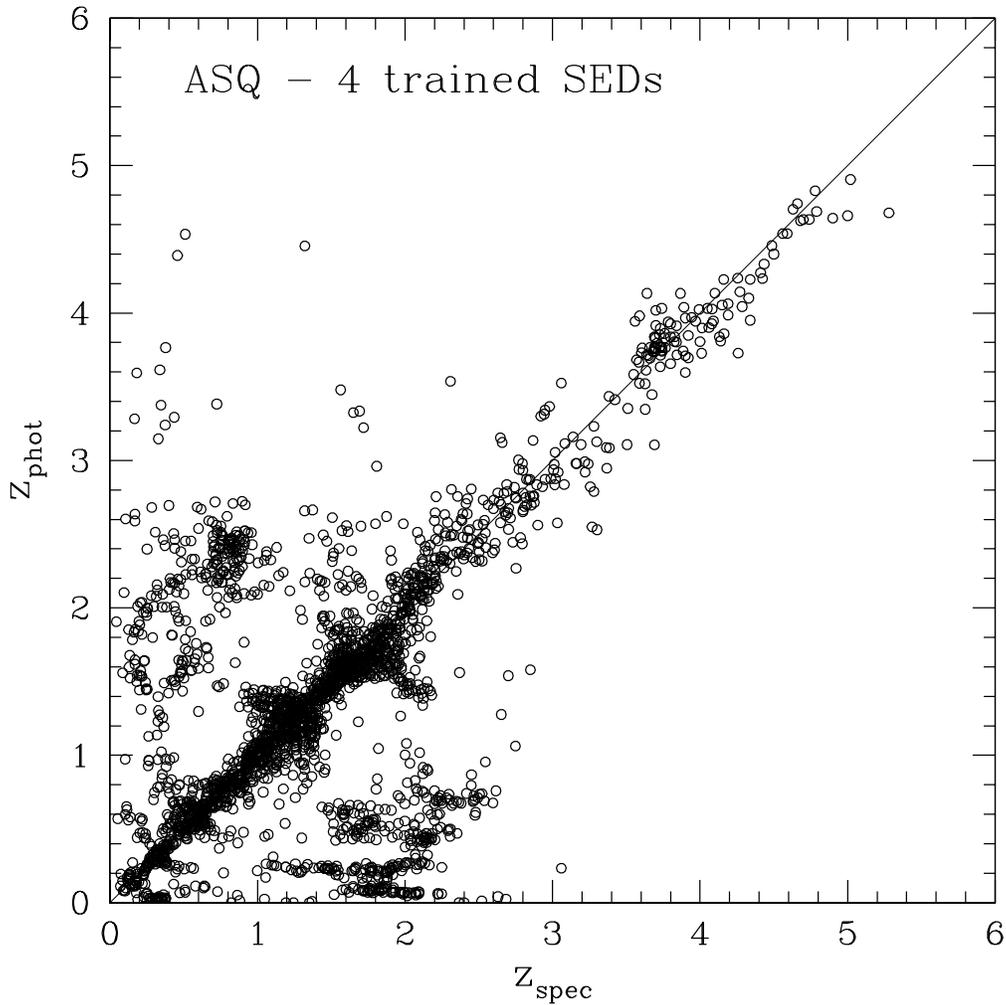}
\caption{The ASQ training yields templates, that provide reliable redshift
estimates. Here we plot the results using four reconstructed templates
($\Delta_{\rm all}=0.77$, $\Delta_{0.3}=0.120$). The degeneracy is clearly
present in the plot just like in the NN estimates.
\label{fig:zz4}}
\end{figure}

\begin{figure}
\epsscale{0.99}
\plotone{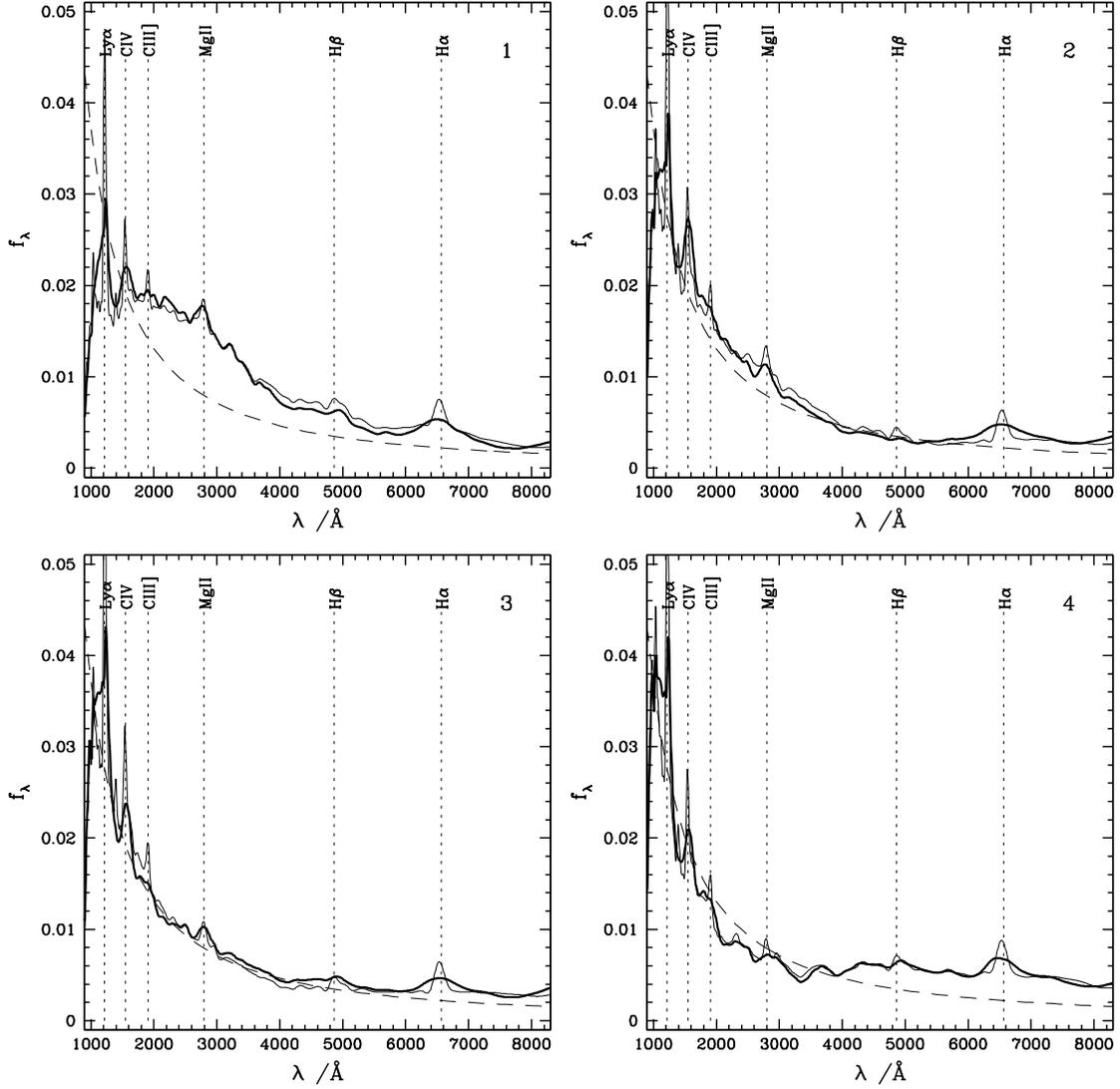}
\caption{Four templates reconstructed from a constant initial SED (thick
line) and from the composite SDSS SED (thin line). Despite the different
initialization, the reconstruction algorithm converges to similar sets of
templates.  The strongest known quasar lines (markers) appear in the
reconstructed spectra.  The same power law curve ($f_{\lambda} \propto
\lambda^{-1.5}$) is plotted in all panels with dashed line to guide the eye
to see the differences.
\label{fig:foursed}}
\end{figure}

Compared to Figure~\ref{fig:zz1}, in Figure~\ref{fig:zz4}, the number of
outliers at $z_{\rm phot} \approx 0$ and 3.5 is greatly reduced and the
redshift estimation improves by increasing the number of templates from one
to four. Using the four ASQ templates gives $\Delta_{\rm all} = 0.77$ and
$\Delta_{0.3} = 0.120$, which is still above the results of empirical
methods. The algorithm can be started from a single constant function or from
the empirical composite SED. Both starting points lead to templates with
approximately the same continuum features (see Figure~\ref{fig:foursed}). The
only difference is that the emission lines remain sharper in the latter
case. The match between the two sets is even more interesting if we remember
that the two template libraries were created using significantly different
initial spectra. This shows that the spectral classification of quasars into
these spectral types is robust.

\begin{figure}
\epsscale{0.85}
\plotone{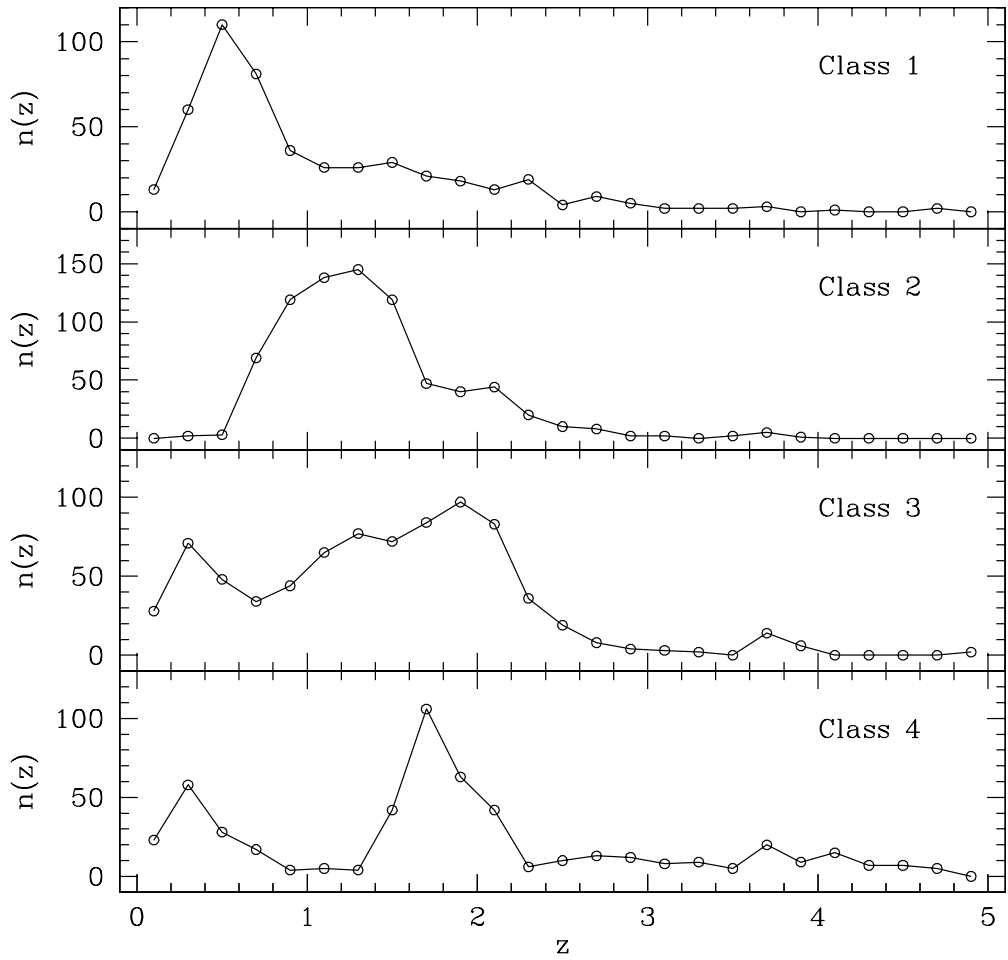}
\caption{Redshift distribution of  objects in  the four ASQ template
classes.
\label{fig:classes}}
\end{figure}

During the photometric redshift estimation with the template fitting method
each object is assigned uniquely to a spectral template derived earlier by
the ASQ algorithm. To test if there is any correlation between the redshift
of the objects and their spectral class, we plot the redshift distributions
of the 4 ASQ classes in Figure~\ref{fig:classes}. We see a relation between
type and redshift, which may have several reasons. Among other
possibilities, it can be related to the different mean absolute luminosity of
the classes, different mixture of BAL and other quasar types, sampling issues
or may be a sign of QSO spectral evolution.  It is also important to note
that at different redshifts the photometric passbands sample different
portions of the templates, and this can be accounted for having low redshift
objects in class 3 and 4, where the highest redshift quasars are.  Detailed
analysis of this is subject of a future work.

\begin{figure}
\epsscale{0.99}
\plotone{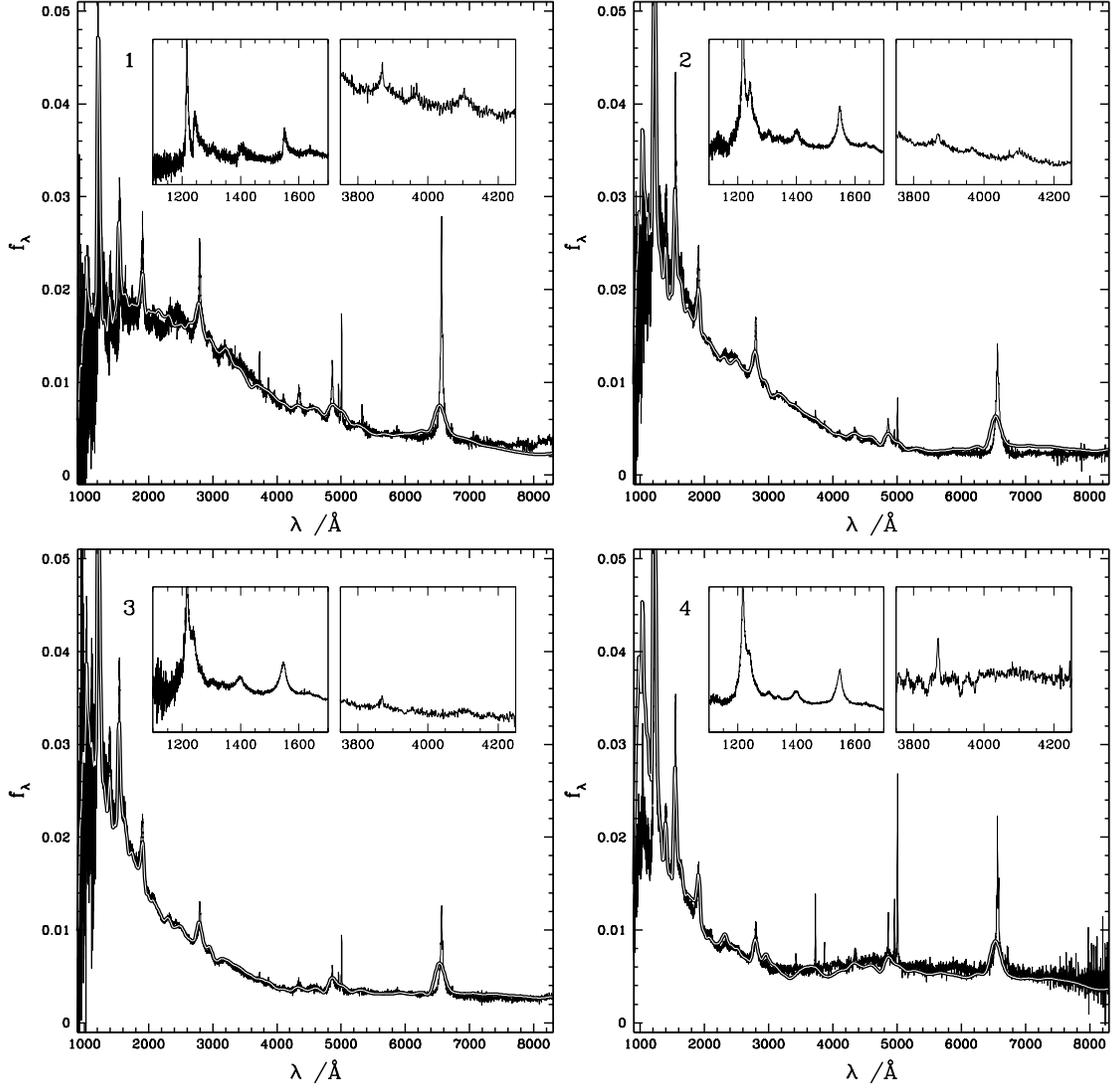}
\caption{In this figure, we compare the ASQ templates (white lines) to
composites created for the corresponding classes using SDSS spectra (thin
lines). See discussion in text. The insets zoom on the Ly\,$\alpha$ to
C\,{\sc iv} region and the Ca\,{\sc ii} H and K lines.
\label{fig:fourcomp}}
\end{figure}

We created composites for each ASQ spectral class using SDSS
spectrophotometric observations. In Figure~\ref{fig:fourcomp}, we compare the
resulting composites to the 4 ASQ templates. The order of the templates here
and also in Figure~\ref{fig:foursed} corresponds to the increasing mean
redshift value of quasars in the classes. Comparing the continuum shape of
the ASQ SEDs (the same power law $f_{\lambda} \propto \lambda^{-1.5}$ is
plotted with dashed lines to help the comparison) suggests a trend with
redshift.

Aside from the apparent redshift dependence of each class, there are
at least two other distinct types of quasars which seem to be at least
partially segregated into the different classes: broad absorption line
quasars (BALQSOs), and low-redshift, low-luminosity quasars with
host galaxy contamination.

Broad absorption features are evident in the class 1 composite spectrum,
and also to a lesser extent in the class 2 composite spectrum.  The
BALQSOs in the quasar sample have been identified, and the majority
($71\%$) fall into either class 1 or class 2.  The fraction of quasars
that have BAL features in their spectra for each class (1,2,3,4) is
$11\%, 6\%, 1.9\%$ and $3\%$ respectively.  The colors of BALQSOs tend
to be redder than non-BAL quasars (Menou \etal\ 2001;
Sprayberry \& Foltz 1992; Brotherton \etal\ 2000;
Weymann \etal\ 1991), which probably accounts for their unequal
separation here into the different classes.

The evidence for host galaxy contamination comes from the presence of stellar
absorption lines in the composite spectra.  For example, the Ca\,{\sc ii} H
and K lines (insets in Figure~\ref{fig:fourcomp}) are clearly present in the
class 4 composite, and weakly present in the class 3 composite.  There is no
evidence for them in either the class 1 or class 2 composites.  Host galaxy
contamination is likely to be a significant contributor to the reddening of
quasars and other AGN at wavelengths beyond about $5000${\AA}
\citep{vanden01}.  For redshifts less than 1, the quasars in classes 3 and 4
are significantly less luminous than their counterparts in classes 1 and 2,
and the quasars with the lowest redshifts tend to belong to classes 3 and 4.
This points to host galaxy contamination in the spectral light of the
low-redshift quasars in classes 3 and 4, and is likely the cause of their
being preferentially separated into those classes.

\begin{figure}
\epsscale{0.99}
\plottwo{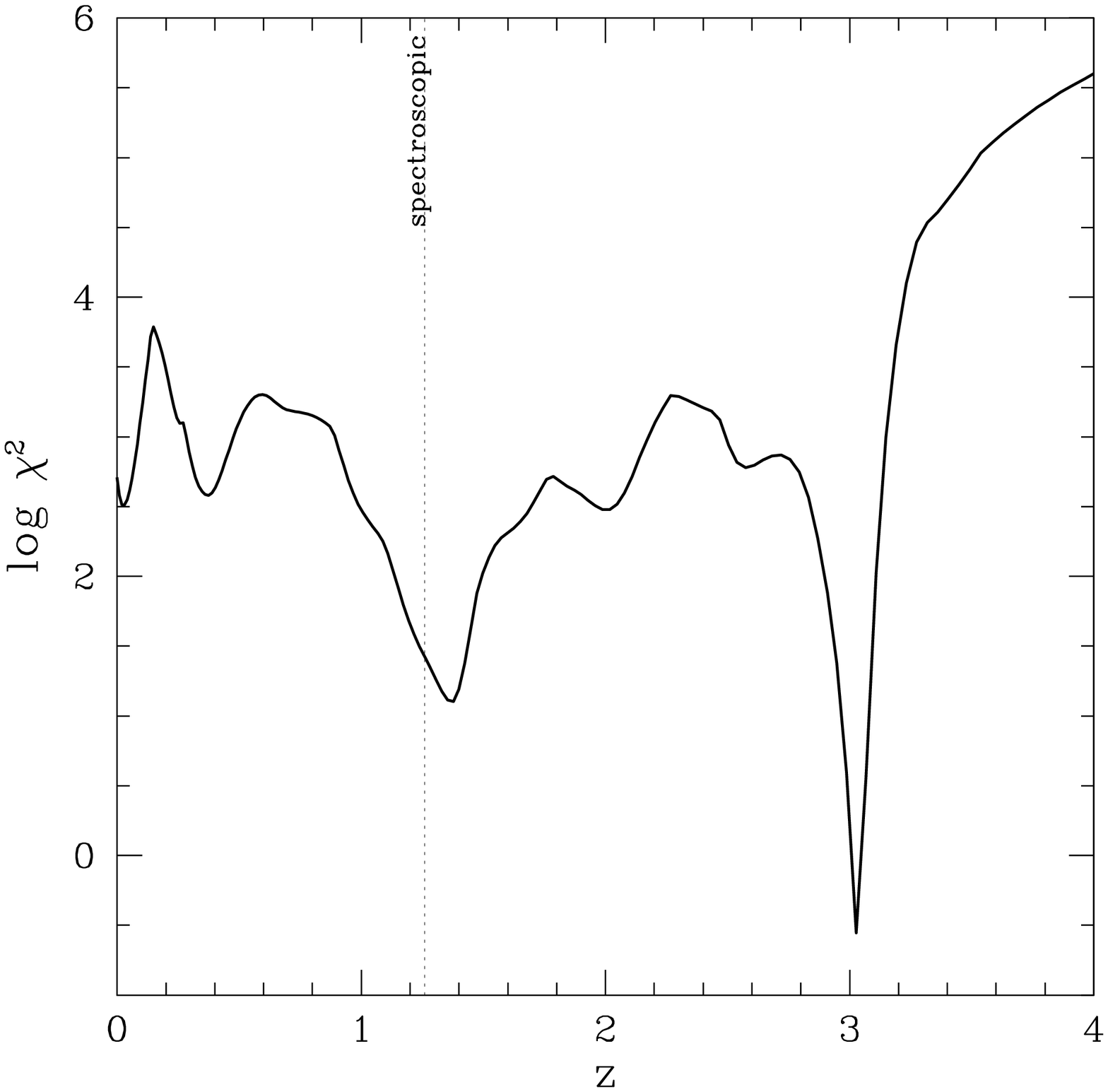}{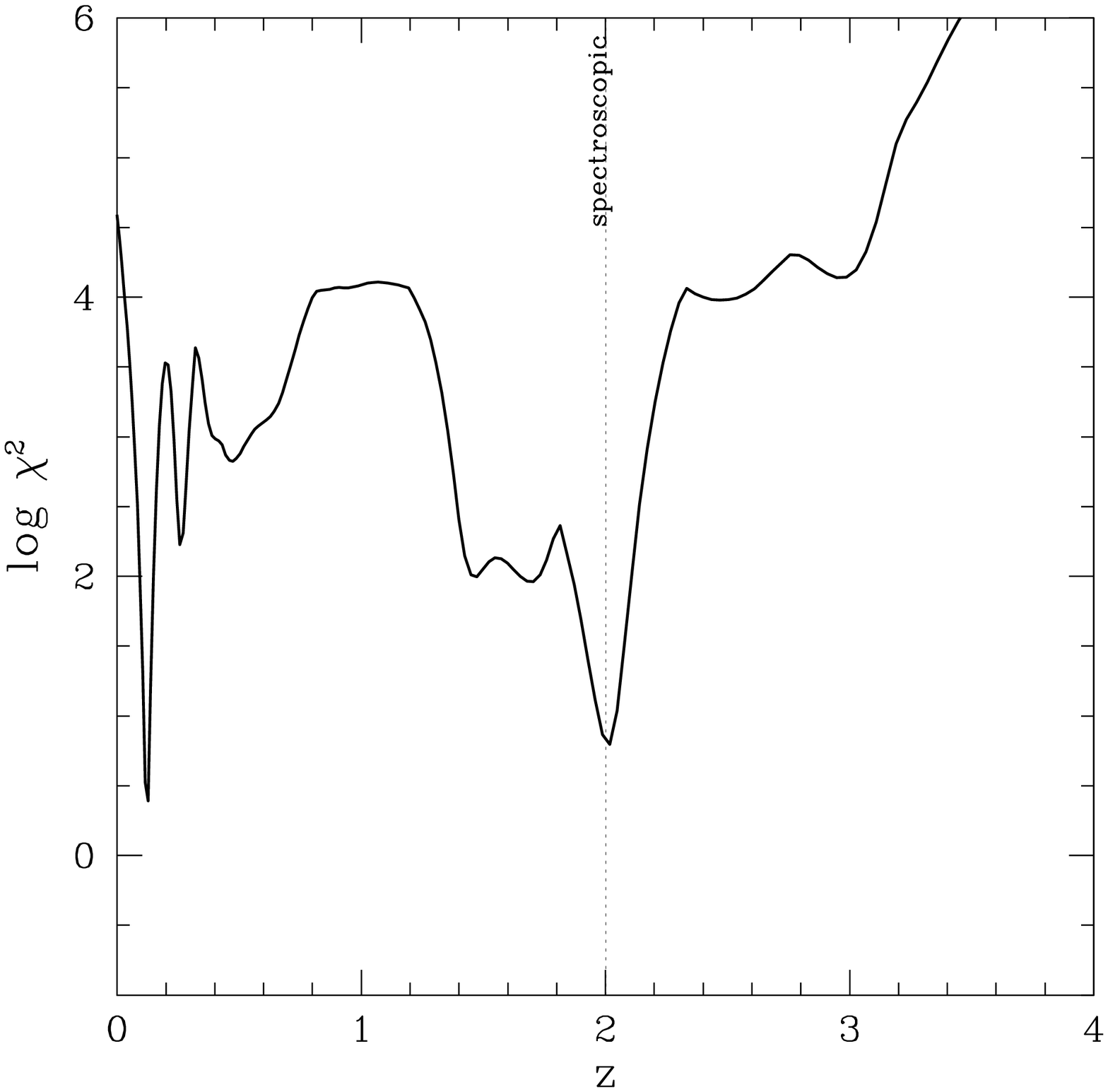}
\caption{The $\chi^2$ curve describes the agreement of the most likely
template and the photometric measurements at an assumed redshift.
The figure illustrates the degeneracy in photometry by plotting $\chi^2$
vs.\ $z$, where more minima are present. 
For outliers the algorithm gives a false redshift value at the global
minimum, but there is also a significant local minimum at the true
spectroscopic redshift (vertical dotted lines).
\label{fig:chi}}
\end{figure}

Now we return to the problem of the photometric degeneracy (i.e.\ quasars
with the same photometric colors can have significantly different redshifts).
Signs of the photometric degeneracy of quasars is visible in the photometric
redshift prediction process.  The $\chi^2$, roughly the weighted distance of
the observed and estimated flux as a function of redshift $z$ has multiple
minima due to the degeneracy in photometry.  Two outliers from the
photometric redshift relation are shown in Figure~\ref{fig:chi}.  Both have
significant minima at the spectroscopic redshift.

\begin{figure}
\epsscale{0.85}
\plotone{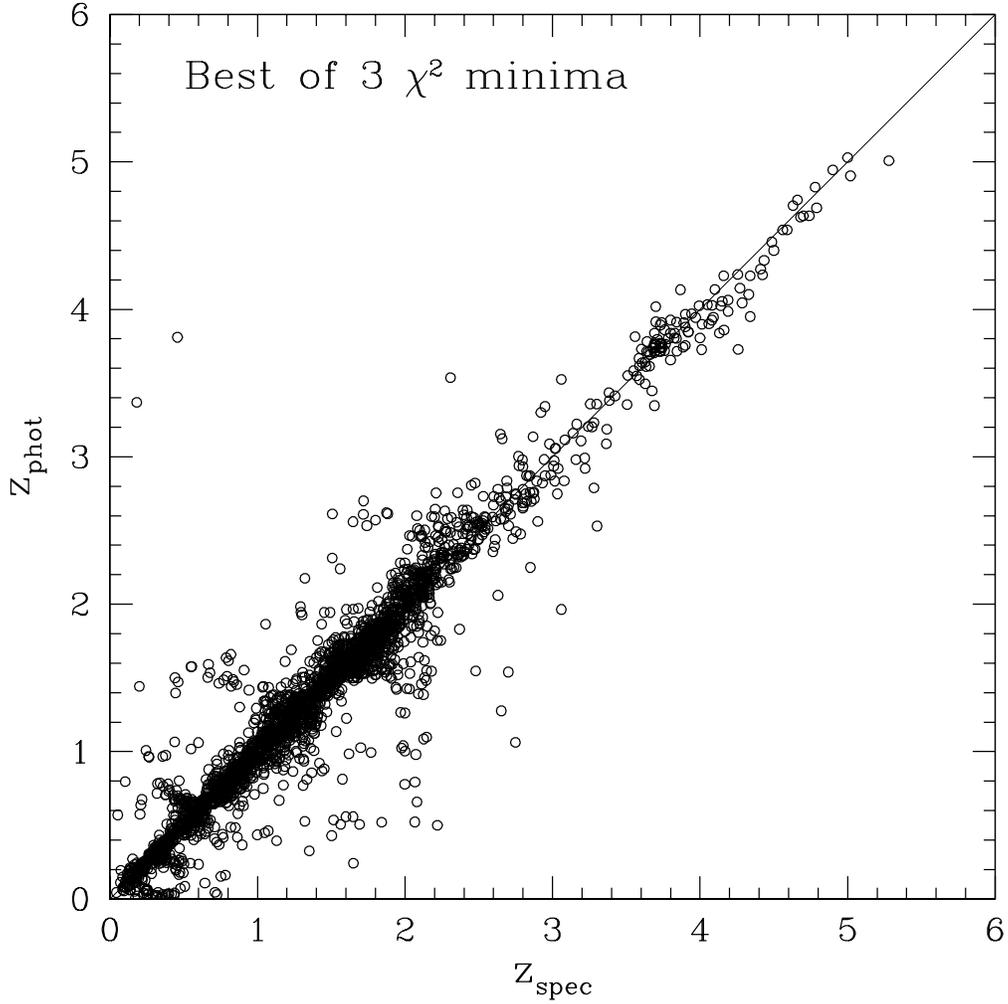}
\caption{The same ASQ templates (see Figure~\ref{fig:zz4}) could provide much
better redshift estimates, if the photometric degeneracy could be broken
e.g.\ by applying some prior. The figure shows a somewhat idealized case: we
use the closest $z_{\rm phot}$ value to $z_{\rm spec}$ from the three
smallest $\chi^2$ minima ($\Delta_{\rm all}=0.238$ and $\Delta_{0.3}=0.104$).
\label{fig:zz4best}}
\end{figure}

In Figure~\ref{fig:zz4best} we plot what one would get if one could break
this degeneracy by using further observations (e.g.\ UV) or through the use
of priors that can crudely constrain the possible redshift of an object. The
photometric redshift estimation algorithm can find all the local minima of
the $\chi^2$ curve, hence we could have more redshift candidates for an
object to choose from.  Given the best 3 local minima for each object, we
selected the one having the closest value to the spectroscopic redshift; this
somewhat idealized case is plotted in the figure.

\section{Conclusions} \label{sec:disc}

The success of our attempt to determine redshifts of quasars from SDSS
multicolor photometry, along with the companion \citet{richards01b} paper, is
encouraging. We can obtain redshift estimates for 70\% of the QSOs with
accuracy $\Delta z_{\rm rms} \approx 0.1$. The method discussed in this
paper is not only able to obtain redshifts, but also to estimate luminosities
and SEDs for QSO candidates without spectroscopy.  Photometric redshifts are
ideally suited for many astrophysical and astronomical studies, such as the
luminosity function or gravitational lensing.

Our spectral template reconstruction appears to be very robust, the algorithm
yields similar sets of spectral templates whether started from a constant
function or from a composite spectrum of spectrophotometric measurements.
Individual spectral lines have significant effects on quasar colors. These
narrow (compared to photometric bands) features contribute to photometric
observations sufficiently that they can be recovered by our reconstruction
technique.
It is quite reassuring that the spectral templates reconstructed from
the very low resolution photometric data show the well known emission
lines of quasars.  This high resolution is possible because of the
large number of SDSS objects used in the template reconstruction
process. Detailed analysis of the templates can give interesting
information on quasar spectral features.  
We found that the use of multiple templates in the redshift estimation
improves the photometric redshift relation and the resulting spectral types
correlate with redshift.

The significance of the spectral lines in broadband photometry also means
that potential errors in the boundary of the assumed passbands may result in
more serious problems in the photometric redshift estimation than is the case
for galaxies where broadband spectral features dominate. If the edge of a
photometric filter were off just by the width of a spectral line, then the
colors, hence the estimated photometric redshifts, would change
substantially.  The fact that the reconstructed spectral lines match well the
location of the actual features is an indication that the filter curves are
indeed accurately calibrated.
For surveys with both photometric and spectroscopic observations (like SDSS),
our method is well suited to cross-check calibration (e.g.~to photometrically
calibrate spectroscopic observations) or may be used to detect
systematic drifts in annual, seasonal, etc.\ data sets.

The template reconstruction method can be extended to derive photometrically
calibrated SEDs for objects fainter than the limit where spectrophotometric
observations are available.  Rough photometric redshift estimates can be also
used in the algorithm as long as there are no systematic deviations, only
scatter around the true value.  An iterative process could be developed to
fine tune the templates.  It could be especially useful for faint high
redshift QSOs, where spectroscopy is not available, but the Ly\,$\alpha$
attenuation could give a fairly good redshift estimate.

\acknowledgments

  The Sloan Digital Sky Survey (SDSS) is a joint project of The University of
Chicago, Fermilab, the Institute for Advanced Study, the Japan Participation
Group, The Johns Hopkins University, the Max-Planck-Institute for Astronomy
(MPIA), the Max-Planck-Institute for Astrophysics (MPA), New Mexico State
University, Princeton University, the United States Naval Observatory, and
the University of Washington. Apache Point Observatory, site of the SDSS
telescopes, is operated by the Astrophysical Research Consortium (ARC).
  Funding for the project has been provided by the Alfred P. Sloan Foundation,
the SDSS member institutions, the National Aeronautics and Space
Administration, the National Science Foundation, the U.S. Department of
Energy, the Japanese Monbukagakusho, and the Max Planck Society. The SDSS Web
site is http://www.sdss.org/.
  I.C. and T.B.  acknowledge partial support from the  MTA-NSF grant no.\ 124
and the  Hungarian National Scientific Research Foundation  (OTKA) grant no.\
T030836.  A.S.  acknowledges support  from NSF (AST9802980)  and a  NASA LTSA
(NAG53503). A.J.C.   acknowledges partial support from  NSF grants AST0096060
and AST9984924 and an NASA LTSA grant NAG5 8546.

\newpage

\newpage

\begin{thebibliography}{}

\bibitem[Ben\'{\i}tez(2000)]{benitez00} Ben\'{\i}tez, N., 2000, \apj, 536,
        571

\bibitem[Brunner, Connolly \& Szalay(1999)]{brunner99} Brunner, R.J., Connolly,
	A.J., \& Szalay, A.S., 1999, \apj, 516, 563

\bibitem[Budav\'ari \etal (1999)]{budavari99} Budav\'ari, T., Szalay, A.S.,
        Connolly, A.J., Csabai, I., \& Dickinson, M.E., 1999, in 
        {\em Photometric Redshifts and High Redshift Galaxies,}  eds.\ R.J.\
        Weymann, L.J.\ Storrie--Lombardi, M.\ Sawicki, \& R.\ Brunner, (San
        Francisco:  ASP), 19

\bibitem[Budav\'ari \etal (2000)]{budavari00} Budav\'ari, T., Szalay, A.S.,
        Connolly, A.J., Csabai, I., \& Dickinson, M.E., 2000, \aj, 120, 1588

\bibitem[Budav\'ari \etal (2001)]{budavari01} Budav\'ari, T., \etal, 2001, in
	preparation 

\bibitem[Connolly \etal (1995a)]{connolly95a} Connolly, A.J., Csabai, I.,
        Szalay, A.S., Koo, D.C., Kron, R.G., \& Munn, J.A., 1995a, \aj, 110,
        2655

\bibitem[Connolly \etal (1999)]{connolly99} Connolly, A.J., Budav\'ari, T.,
        Szalay, A.S., Csabai, I., \& Brunner, R.J., 1999, in 
        {\it Photometric Redshifts and High Redshift Galaxies,}  eds.\ R.J.\
        Weymann, L.J.\ Storrie--Lombardi, M.\ Sawicki, \& R.\ Brunner, (San
        Francisco:  ASP), 13

\bibitem[Csabai \etal (2000)]{csabai00} Csabai, I., Connolly, A.J., Szalay,
        A.S., \& Budav\'{a}ri, T., 2000, \aj, 119, 69

\bibitem[Fern\'andez-Soto \etal (1999)]{soto99} Fern\'{a}ndez-Soto, A.,
        Lanzetta, K.M., \& Yahil, A., 1999, \apj, 513, 34

\bibitem[Francis \etal (1991)]{francis91} Francis, P.J., Hewett, P.C., Foltz,
	C.B., Chaffee, F.H., Weymann, R.J., Morris, S.L., 1991, \apj, 373, 465

\bibitem[Fukugita \etal (1996)]{fukugita} Fukugita, M., Ichikawa, T., Gunn,
	J.E., Doi, M., Shimasaku, K. \& Schneider, D.P. 1996, \aj, 111, 1748

\bibitem[Gwyn \& Hartwick(1996)]{gwyn96} Gwyn, S.D.J., \& Hartwick, F.D.A.,
        1996, \apj, 468, L77

\bibitem[Kinney \etal (1996)]{kinney96} Kinney, A.L., Calzetti, D., Bohlin,
        R.C., McQuade, K., Storchi-Bergman, T., \& Schmitt, H.R., 1996, \apj,
        467, 38 

\bibitem[Kohonen(1995)]{kohonen} Kohonen, T., 1995,  Self-Organizing Maps,
	Springer Series in Information Sciences

\bibitem[Koo(1985)]{koo85} Koo, D.C., 1985, \aj, 90, 148

\bibitem[Lupton \etal (2001)]{lupton} Lupton, R.H., \etal, 2001, in preparation

\bibitem[Madau(1995)]{madau95} Madau, P., 1995, \apj, 441, 18

\bibitem[Richards \etal (2001a)]{richards01a} Richards, G.T.,
	\etal, 2001a, \aj, 121, in press

\bibitem[Richards \etal (2001b)]{richards01b} Richards, G.T., \etal, 2001b,
	this volume

\bibitem[Sawicki, Lin \& Yee(1997)]{sawicki97} Sawicki, M.J., Lin, H., \& Yee,
        H.K.C, 1997, \aj, 113, 1 

\bibitem[SDSS; York \etal (2000)]{york} York, D.G., \etal, 2000, AJ, 120,
	1579

\bibitem[Vanden Berk \etal (2001)]{vanden01} Vanden Berk, D.E., \etal, 2001,
	in preparation

\bibitem[Wang, Bahcall \& Turner(1998)]{wang98} Wang, Y., Bahcall, N., \&
        Turner, E.L., 1998, \aj, 116, 2081

\bibitem[Weymann \etal (1999)]{weymann99} Weymann, R.J., 
        Storrie--Lombardi, L.J., Sawicki, M., \& Brunner, R., (editors), 1999,
        {\em Photometric Redshifts and High--Redshift Galaxies} (San
        Francisco:  ASP)

\end{thebibliography}
\end{document}